\documentclass[journal]{IEEEtran}
\usepackage{amsmath,amsfonts}
\usepackage{algpseudocode}
\usepackage{amssymb}
\usepackage{algorithm}
\usepackage{array}
\usepackage[caption=false,font=normalsize,labelfont=sf,textfont=sf]{subfig}
\usepackage{textcomp}
\usepackage{stfloats}
\usepackage{url}
\usepackage{verbatim}
\usepackage{graphicx}
\usepackage{cite}

\usepackage{threeparttable}
\usepackage{multirow}

\newcommand{\figref}[1]{Fig.~\ref{#1}}                      
\newcommand{\tabref}[1]{Table~\ref{#1}}                      
\newcommand{\eqnref}[1]{Eq. \eqref{#1}}                     

\begin{document}

\title{Versatile Volumetric Medical Image Coding for 
Human-Machine Vision}

\author{Jietao Chen, Weijie Chen, Qianjian Xing and Feng Yu,~\IEEEmembership{Member,~IEEE}

\thanks{Corresponding author: Qianjian Xing, Feng Yu.}
\thanks{Jietao Chen and Feng Yu are with the College
of Biomedical Engineering \& Instrument Science, Zhejiang University,
Zhejiang, 310063, China. (E-mails: jt\_chen@zju.edu.cn; cwjzzl@zju.edu.cn; xingqianjian@zju.edu.cn; osfengyu@zju.edu.cn).}
}

\maketitle

\begin{abstract}
Neural image compression (NIC) has received considerable attention due to its significant advantages in feature representation and data optimization. However, most existing NIC methods for volumetric medical images focus solely on improving human-oriented perception. For these methods, data need to be decoded back to pixels for downstream machine learning analytics, which is a process that lowers the efficiency of diagnosis and treatment in modern digital healthcare scenarios. In this paper, we propose a Versatile Volumetric Medical Image Coding (VVMIC) framework for both human and machine vision, enabling various analytics of coded representations directly without decoding them into pixels. Considering the specific three-dimensional structure distinguished from natural frame images, a Versatile Volumetric Autoencoder (VVAE) module is crafted to learn the inter-slice latent representations to enhance the expressiveness of the current-slice latent representations, and to produce intermediate decoding features for downstream reconstruction and segmentation tasks. To further improve coding performance, a multi-dimensional context model is assembled by aggregating the inter-slice latent context with the spatial-channel context and the hierarchical hypercontext. Experimental results show that our VVMIC framework maintains high-quality image reconstruction for human vision while achieving accurate segmentation results for machine-vision tasks compared to a number of reported traditional and neural methods.
\end{abstract}

\begin{IEEEkeywords}
Neural networks, Volumetric image compression, Human-machine vision, Medical image segmentation.
\end{IEEEkeywords}

\section{Introduction}
\IEEEPARstart{E}{xtensive} utilization of medical imaging technology, including CT and MRI, for clinical auxiliary diagnosis is well-recognized. As digital imaging technology continues to advance and medical devices achieve higher resolutions, the size and bit depth of digital medical images are progressively expanding, posing challenges for storage and transmission. Meanwhile, machine learning has significantly pushed forward the frontier of intelligent analytics such as automatic segmentation\cite{xing2024hybrid, xie2022dmcgnet, alam2022multi} and classification\cite{panayides2020ai, behera2022brain, lao2021regression}. However, for most existing image compression methods, the coded content typically needs to be fully decoded back to pixels before performing downstream machine-vision analysis, which is no longer efficient\cite{liu2023icmh}. To address it, JPEG-AI\cite{jpegai} standard first proposed the concept of image coding for human-machine vision, enabling analyzing coded representations directly without decoding them into pixels. Keeping this concept in mind, this paper aims to propose a medical image coding framework focused on human-machine vision for the demand of modern digital healthcare scenarios.
\begin{figure}[!t]
\centering
\includegraphics[width=3.4in]{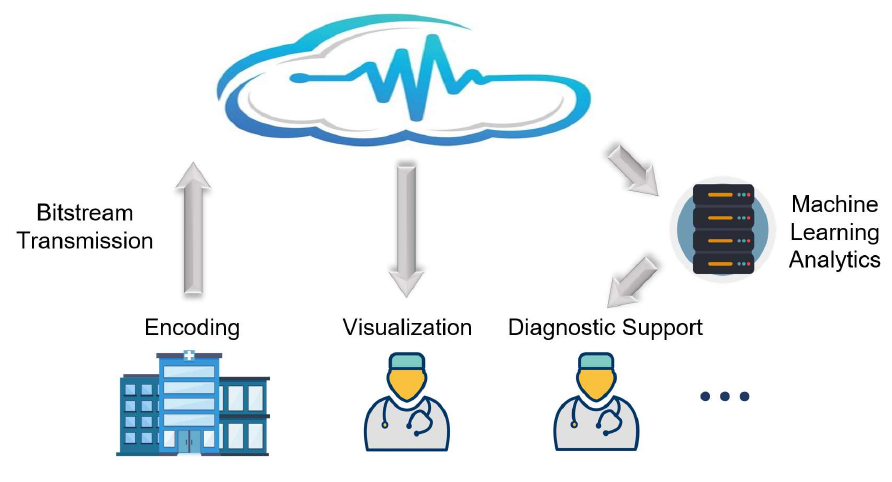}
\caption{An example of medical image compression for human-machine vision in the medical cloud.}
\label{fig:abstract}
\end{figure}

Currently, many medical imaging systems adhere to the guidelines established by the Digital Imaging and Communications in Medicine (DICOM) organization, which specify the use of traditional image coding standards, including PNG\cite{PNG}, JP2K\cite{JP2K}, JP3D\cite{JP3D}, etc. In addition, next-generation image compression algorithms, such as FLIF\cite{flif} and JPEG-XL\cite{jpegxl}, have emerged rapidly in recent years. The traditional video coding standards like HEVC\cite{HEVC}, and VVC\cite{vvc} are also candidates for volumetric medical image compression algorithms. Furthermore, neural image compression (NIC) methods have also achieved tremendous success in the field of image compression\cite{balle2016end}. For example, Ballé \textit{et al.}\cite{balle2016end, hyperprior} first adopted VAE-based architecture for neural image compression. Minnen \textit{et al.}\cite{minnen_joint_2018} introduced a joint auto-regressive based entropy model, while Cheng \textit{et al.}\cite{cheng2020} further improved the auto-regressive based entropy model by adopting the Gaussian mixture model for context. Additionally, some works\cite{elic, contextformer, entroformer, stf, mlic, liu2023mixed} utilize attention mechanism \cite{attention} to enhance the reconstructed image quality.

Although the aforementioned traditional and neural methods achieve promising compression performance, most of them are only optimized for human perception. Taking a scenario of medical clouds as an example, illustrated in \figref{fig:abstract}, after one client transmits the compressed images to the server, the bitstream must be restored to pixels before downstream analysis can be performed. What's worse, using reconstructed pixels from lossy compression as inputs for downstream analysis models may even lead to performance degradation. Another kind of image compression methods like\cite{torfason2018towards, song2021variable, liu2021semantics, le2021image, liu2019machine} aim at compression for machines. For example, Liu \textit{et al.}\cite{liu2019machine} proposed a machine vision-oriented approach based on learning-based 3D-DWT, achieving better medical image segmentation performance compared to JP2K at the same compression ratio. However, purely machine vision-oriented methods have certain limitations in the cloud scenario; apart from the master client that uploads the original image's compressed bitstream to the server, other clients cannot restore images tailored for human vision. Therefore, proposing an effective NIC method for both human and machine vision is emergent. 

This paper introduces a Versatile Volumetric Medical Image Coding framework (VVMIC) for human and machine vision, referring to the three standards of JPEG-AI\cite{jpegai}, i.e., it should: 1) use a single bitstream for both human and machine vision; 2) achieve high compression efficiency for image reconstruction and satisfactory performance for human and machine vision tasks; 3) enable directly performing human and machine vision algorithms without decoding the coded representations into pixels.

In detail, we propose a Versatile Volumetric Autoencoder (VVAE) module to learn the inter-slice latent representations for enhancing the expressiveness of the current-slice latent representations, and produce intermediate decoding features for downstream reconstruction and segmentation tasks. To this end, a recurrent based compression loop is leveraged to load and update the inter-slice latent representations from the auxiliary feature buffer. An inter-slice analysis transform module and an inter-slice synthesis transform module are employed to extract multi-scale auxiliary latent representations, which are then fed into the current image encoder and decoder networks. we also introduce a multi-dimensional context entropy model, which aggregates local spatial context, channel-wise context, inter-slice auxiliary context, and hierarchical hypercontext, thereby improving the coding performance of latent representations. For human vision-oriented reconstruction tasks, we feed intermediate latent decoding features into an image reconstruction network to recover the pixels, while for machine vision-oriented downstream tasks, we input current-slice latent decoding features into a commonly used segmentation network to obtain segmentation masks. 

Our contributions are summarized as follows: 
\begin{itemize}
    \item To the best of our knowledge, it’s the first end-to-end volumetric medical image coding framework towards both human and machine vision, facilitating the use of learning-based compression methods in digital healthcare scenarios like medical clouds.
    \item We introduce a novel latent representation extraction model called the Versatile Volumetric Autoencoding (VVAE), which leverages multi-scale latent features from previous slices to enhance the expressiveness of the current slice features.
    \item We propose a multi-dimensional context model by aggregating inter-slice latent context with spatial-channel context and hierarchical hypercontext, resulting in enhanced coding performance.
    \item We perform extensive experiments on four volumetric image datasets, where we achieve better human-vision reconstructions than traditional compression methods like JPEG-XL, HEVC, and VVC while improving the downstream segmentation results for machine vision.
\end{itemize}

\section{Related Works}
\subsection{Neural Lossy Image Compression}
With the development of deep learning, neural image compression methods have emerged as a promising research direction. 

The first end-to-end optimized image compression network with a learned image encoder and decoder was proposed by Ballé \textit{et al.}\cite{balle2016end}, which can be interpreted as VAEs\cite{autoencoder} based on transform coding\cite{theoretical}. The next year, Ballé \textit{et al.}\cite{hyperprior} improved upon this\cite{balle2016end} by introducing the learned hyperprior network to capture hierarchical dependencies in the latent representation. Minnen \textit{et al.}\cite{minnen_joint_2018} also contributed to this field by adding an autoregressive context module to improve coding performance. Later Cheng \textit{et al.}\cite{cheng2020} utilized attention-based modules and discretized Gaussian Mixture Model to enhance the image compression network. Since then, there have also been many recently learned compression methods that improved in the aspects of transform (network architecture)\cite{ma2020end, zhu2022transformer, liu2023mixed} and entropy model\cite{elic, contextformer, entroformer, mlic}. In detail, He \textit{et al.}\cite{checkerboard} proposed to separate the latents into anchors and non-anchors and to adopt a checkerboard convolution, solving the slow decoding issue caused by the serial autoregressive convolution. To the same end, Minnen \textit{et al.}\cite{channelwise} introduced a channel-wise context prior between channels, and then He \textit{et al.}\cite{elic} improved it through unevenly grouping channel slices.

As for volumetric medical image compression, Gao \textit{et al.}\cite{gao_volumetric_2020} first used a learning-based method to compress 3D images\cite{xue_aiwave_2023}. Xue \textit{et al.}\cite{xue_aiwave_2023} proposed a 3D trained affine wavelet-like transform method called aiWave, which can be adaptive to different data and be feasible to handle various local
contexts. Based on it, Xue \textit{et al.} also introduced the aiWave-lite\cite{xue2024lightweight} and the extended version for 4D datasets\cite{xue_dbvc_2023}.

Although the reconstructed images provided by the above-mentioned methods can be directly used for downstream machine vision tasks, they mainly focus on compressing images for human vision without considering the specific requirements of machine vision tasks. This highlights the need to design a new compression method that caters to both human and machine vision\cite{liu2023icmh}.

\subsection{Compression for human-machine vision}
The massive amount of image and video streaming generated every day poses great challenges in compression, transmission, and analysis. It remains an open problem on how to perform efficient feature compression that meets the different perception requirements of both humans and machines in various computer vision tasks. 

The scope of JPEG-AI\cite{jpegai} is the creation of a learning-based image coding standard offering a single-stream, compact compressed domain representation, targeting both human visualizations, with significant compression efficiency improvement over image coding standards. In addition, the MPEG VCM\cite{vcm} working group from ISO/ITU is developing advanced feature compression technologies, which extract compact representations for both visual features and videos as bridges to connect machine vision and human vision tasks.

Responding to the calls from these groups, recently many researchers have explored neural compression frameworks for human-machine vision. For instance, Hu \textit{et al.}\cite{hu2020towards} encoded critical image structure and color information separately, and then utilized a generative network to recover images for both human vision and facial landmark detection. Choi \textit{et al.}\cite{choi2022scalable} designed the latent space to support scalability from simpler to more complicated tasks of human-machine tasks, achieving more bitrate savings while keeping comparable reconstruction qualities. Wang \textit{et al.}\cite{wang2022human} then explored the human-machine interaction-oriented image coding under the scenarios of the Internet of Things. Liu \textit{et al.}\cite{liu2023icmh} selectively chose the optimal subset of quantized latent features based on a learnable binary mask, to adaptively partition, transmit, reconstruct, and aggregate (PTRA) the latent representations for both machine tasks and human vision. Finally, Sheng \textit{et al.}\cite{sheng2024vnvc} proposed a versatile neural video coding (VNNC) framework for video reconstruction and analytics, in which an intermediate feature was used as a reference in motion compensation and motion estimation.

However, most above-mentioned methods are based on 2D images or videos, which may not be suitable for extracting specific inter-slice redundancy for 3D medical images. Video coding techniques with motion estimation and compensation struggle to handle these images due to the absence of apparent background and moving objects, making it counter-intuitive to treat the inter-slice dimension as a time domain\cite{xue_aiwave_2023}. Additionally, the coding for the machine method mentioned in the literature\cite{liu2019machine} only ensures the performance of medical image segmentation while neglecting the quality of the reconstructed images, making it inadequate to meet the demand of digital healthcare scenarios. 

Therefore, this paper proposes a 3D medical image compression framework focused on human-machine vision and designs an efficient inter-slice feature extraction network named VVAE, which improves the compression of current-slice latent features and enhances the performance of downstream segmentation tasks.

\begin{figure*}[!t]
\centering
\includegraphics[width=7in]{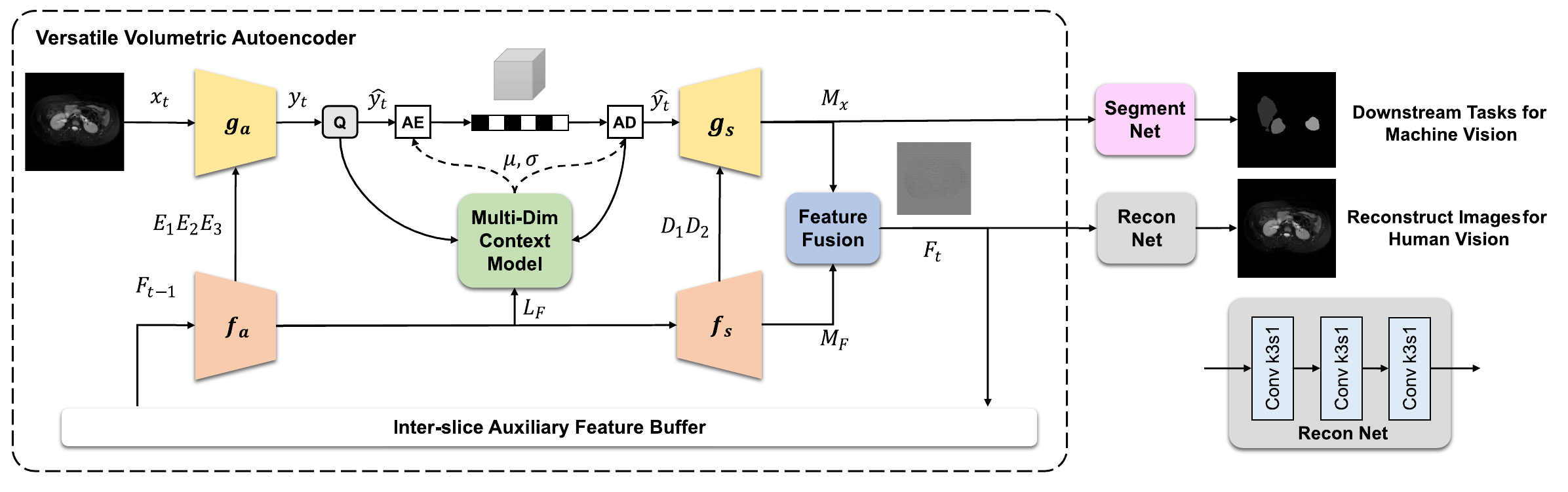}
\caption{The details of our proposed VVMIC framework, which codes volumetric medical images in a recurrent compression loop. The auxiliary analysis module $f_a$ and the synthesis module $f_s$ encode and decode the buffered inter-slice auxiliary features $F_t$, producing multi-scale encoding representations $E_1E_2E_3$, decoding representations $D_1D_2$, and inter-slice context priors $L_F$. Aggregating these features, the current-slice autoencoder network completes the encoding of $x_t$ and finally decodes it into intermediate features $M_x$ and $M_F$, which are then fused to update inter-slice auxiliary feature $F_t$ and fed to downstream tasks for human-machine vision.}
\label{fig:framework}
\end{figure*}

\begin{figure*}[!t]
\centering
\includegraphics[width=7in]{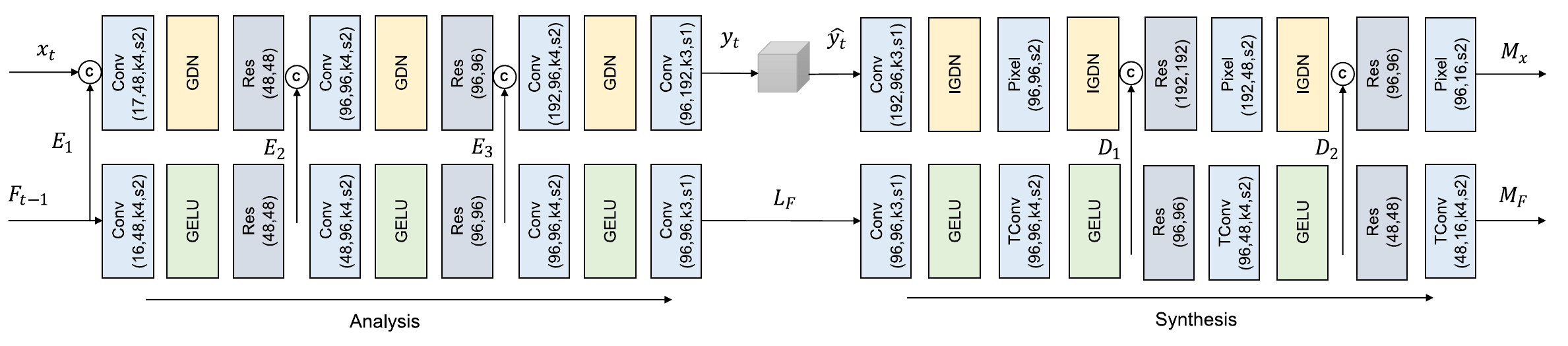}
\caption{The network structure of our versatile volumetric autoencoder (VVAE).}
\label{fig:network}
\end{figure*}

\section{Method}
This section introduces our proposed volumetric medical image compression framework (VVMIC) in detail. First, we give a concise introduction to the autoencoder-based lossy compression method. Based on this, we present our versatile volumetric autoencoder (VVAE) module for learning inter-slice and current-slice latent representations, and our multi-dimensional context model for coding them. Finally, we discuss the strategies for image restoration for human vision and the downstream segmentation task for machine vision.

\subsection{Prerequisites}
According to rate-distortion optimization, given input image $x$, a small bit-rate $R$ will lead to large distortion $D$. To adjust the balance of rate and distortion, $\lambda$ is used to control the optimization process, which can be formulated as:
\begin{equation}
\mathcal{L} = R + \lambda D
\end{equation}

Since Ballé \textit{et al.}\cite{balle2016end} proposed the first end-to-end lossy image compression method, Minnen \textit{et al.}\cite{minnen_joint_2018} introduced a foundational framework that has influenced many subsequent research works\cite{cheng2020, stf, checkerboard, channelwise, elic, mlic}. This kind of framework consists of an analysis encoder $g_a$, a synthesis decoder $g_s$, a quantization $Q$, a hyper analysis encoder $h_a$, a hyper synthesis decoder $h_s$, a context entropy model $g_{ctx}$, and an entropy parameter estimator $g_{ep}$. The principle of R-D optimization is utilizing encoders and decoders to reconstruct the image $\hat{x}$ with low distortion $D$ from the original image $x$, while exploiting hierarchical hyperparameters $\Psi$ and context priors $\Phi$ to keep low rate $R$. The whole process is as follows:
\begin{gather}
y=g_a(x), \hat{y}=Q(y),  \hat{x} =g_s(\hat{y}) 
\label{eqn:analysis_synthesis}                 \\
\Phi = g_{ctx}(\hat{y}), \Psi = h_s(Q(h_a(y))) \\
\Theta_{\mu,\sigma} = g_{ep}(\Phi,\Psi) 
\label{eqn:theta}
\end{gather}
where $y$ and $\hat{y}$ represents the latents and the quantized latents respectively.

Initially, the context information\cite{minnen_joint_2018, cheng2020} was extracted by pixel-cnn-like masked convolution network\cite{pixelcnn}. The spatial context features $\Phi_{sp}$ are obtained by the transform function $g_{sp}$ from observable neighbors $\hat{y}_{<i}$ of each symbol vector at the i-th location:
\begin{align}
\hat{y}_{<i} &= \{ \hat{y}_1,...,\hat{y}_{i-1}\} \\
\Phi_{sp,i} &= g_{sp}(\hat{y}_{<i})
\end{align}
This approach demands symbols $\{ \hat{y}_1,\hat{y}_2,...,\hat{y}_N \}$ to be decoded serially, which critically consumes time and computational resources\cite{elic}. To address this issue, He \textit{et al.}\cite{checkerboard} proposed to separate the latent representation $\hat{y}$ into the anchor part $\hat{y}_{anc}$ and the non-anchor part $\hat{y}_{nonanc}$, taking $\hat{y}_{anc}$ as the context prior of $\hat{y}_{nonanc}$ to achieving two-pass parallel decoding:
\begin{equation}
\hat{y}_{<i}^{anc} = \varnothing, \hat{y}_{<i}^{nonanc} = \hat{y}^{anc}
\label{eqn:ckbd}
\end{equation}

Another scheme to perform parallel backward adaption is to model contexts between channels\cite{channelwise}. $\hat{y}$ is evenly divided to $K$ channel-slices, and the current channel-slice $\hat{y}^k$ is conditioned on previously decoded ones $\hat{y}^{<k}$. Then the context priors of the current channel can be expressed as:
\begin{equation}
\Phi_{ch}^k = g_{ch}^k(\hat{y}^{<k}), k=2,...,K
\end{equation}
where $g_{ch}$ represents the transform function of channel-wise context.

\subsection{Framework}
Notably, within the volume of 3D medical images, a wealth of redundant information exists between slices. This presents a challenge: effectively extracting inter-slice details to serve as prior knowledge for compressing the current image and improving the performance of downstream analytic tasks.

As illustrated in \figref{fig:framework}, our VVMIC framework consists of the versatile volumetric autoencoder (VVAE) module and the downstream task networks. Before compressing the current slice $x_t$, the VVAE module first loads inter-slice auxiliary features as $F_{t-1}$, which contains prior information of previous slices. Multi-scale inter-slice contexts $E_1, E_2, E_3$ and inter-slice latent representations $L_F$ are learned by the auxiliary analysis transform module $f_a$:
\begin{equation}
E_1,E_2,E_3,L_F = f_a(F_{t-1})
\end{equation}
These multi-scale contexts are then channel-wisely concatenated into the current image analysis module $g_a$ to enhance the expressiveness of current-slice representations:
\begin{equation}
y=g_a(x,E_1,E_2,E_3)\\
\end{equation}

In the decoding procedure, the auxiliary synthesis transform module $f_s$ then learns further multi-scale inter-slice contexts $D_1,D_2$ from the inter-slice latent representations $L_F$:
\begin{equation}
D_1,D_2,M_F = f_s(L_F) 
\end{equation}
These $D_1,D_2$ contexts are also concatenated into the current image synthesis module $g_s$:
\begin{equation}
M_x = g_s(\hat{y}, D_1,D_2)
\end{equation}

Then a feature fusion module is employed to aggregate the produced latent decoding features $M_x$ and $M_F$, generating the intermediate features $F_t$. We regard this $F_t$ as the new inter-slice auxiliary features to help compress the next slice $x_{t+1}$. Meanwhile, we also directly feed $F_t$ into the reconstruction network and $M_x$ into the segmentation network for human-machine vision tasks. In this way, the reconstruction process of $x_t$ can be expressed as:
\begin{align}
F_t &= \mathcal{F}(M_x, M_F)\\
\hat{x_t} &=\mathcal{R}(F_t)
\end{align}
where $\mathcal{R}$ represents our reconstruction network, aiming to restore the image $\hat{x}$, and $\mathcal{F}$ represents the feature fusion module. In our framework, we directly employ the $tanh$ gate to aggregate $M_x$ and $M_F$, similar to the approach commonly used in RNN models.

\subsubsection{Analysis and Synthesis Transforms}
To further illustrate the details of our network, \figref{fig:network} shows the specific structure of the analysis and synthesis network layers within our VVAE module. In the analysis module, the inter-slice analysis $f_a$ generates multiple multi-scale inter-slice contexts by stacking convolutional layers, GELU activation layers\cite{gelu}, and residual network layers\cite{residual}. The current-slice analysis module $g_a$ then leverages inter-slice redundancy through concatenation operations. Following previous works\cite{minnen_joint_2018, cheng2020, stf, sheng2024vnvc}, we employ the GDN network\cite{gdn} as the normalization layer to improve non-linearity\cite{mlic}. 

During the decoding process, different from video-based methods using motion vectors to extract inter-slice information, the auxiliary latent representation $L_F$ of our framework does not need to recover from binary files. This mechanism also saves the bit costs to compress the latent representation $L_F$, leading to an improvement in coding performance. For the design of the synthesis module, we employ convolutional layers, IGDN layers, residual network layers, as well as PixelShuffle layers, and transposed convolutional layers for upsampling. The PixelShuffle layer is commonly used as the output layer in image super-resolution networks, helping to preserve the details and textures of images.

\begin{figure}[!t]
\centering
\includegraphics[width=3.4in]{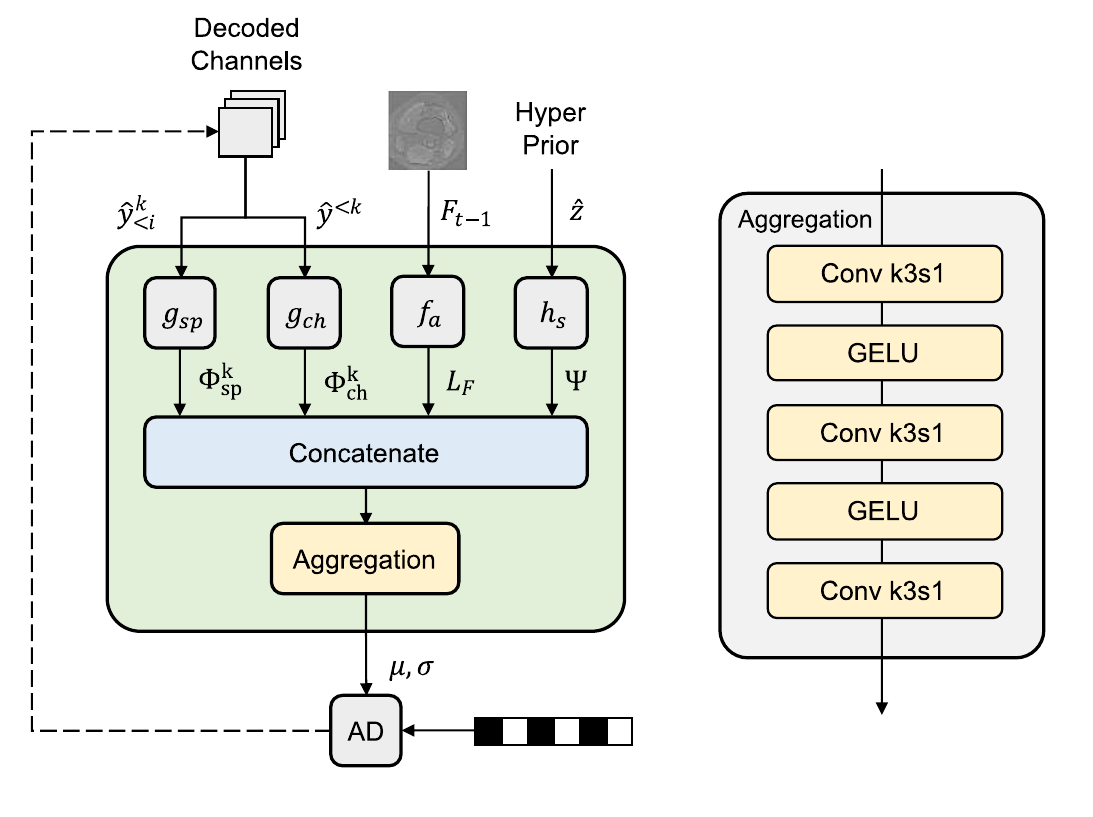}
\caption{The decoding diagram of our Multi-dimension Context Model, which exploits the spatial context priors $\Phi_{sp}^k$ from processed pixels of the current channel $\hat{y}_{<i}^k$, the channel-wise context priors $\Phi_{ch}^k$ from processed channels $\hat{y}^{<k}$, the hierarchical hyper-parameters $\Psi$, and the inter-slice latent representations $L_F$.}
\label{fig:contextmodel}
\end{figure}
\subsubsection{Multi-Dim Context Model}
The context model plays a substantial role in coding performance among VAE-based compression methods\cite{minnen_joint_2018}, which exploits sufficient context priors to reduce the data redundancy of latent representations. In this paper, we propose the Multi-Dimention Context Model as shown in \figref{fig:contextmodel} that aggregates spatial-channel context, hierarchical context, and inter-slice latent priors. The spatial context and channel-conditional models eliminate redundancy along spatial and channel axes respectively\cite{elic, mlic}. Following them\cite{elic}, We combine these two models for better backward-adaptive coding. In detail, we perform $g_{sp}$ transform on the current channel-slice $\hat{y}_{<i}^k$ to learn the spatial context priors $\Phi_{sp}^k$, employing the checkerboard two-pass parallel model\cite{checkerboard} as \eqnref{eqn:ckbd} shows. The $g_{ch}$ transform on the previous channel-slices $\hat{y}^{<k}$ are utilized to learn the channel-wise context priors $\Phi_{ch}^k$.

For volumetric medical images, we assume that the structural similarities between the latent representations of the current and previous slices have not been adequately addressed in the literature. To tackle this, we leverage the inter-slice features as one of the contextual priors for compressing the latent representations of the current slice. Here we directly take the inter-slice analysis transform module $f_a$ as the prior-context extraction function. Based on it,  we concatenate the $L_F$ with the above-mentioned spatial and channel context priors, as well as the hierarchical hyper-parameters $\Psi$. Finally, we employ a entropy parameter estimation module $g_{ep}$ to aggregate these context priors all, where the \eqnref{eqn:theta} can be updated as:
\begin{equation}
\Theta_{\mu,\sigma} = g_{ep}(\Phi_{sp}^k,\Phi_{ch}^k,L_F,\Psi)
\end{equation}

\subsubsection{Reconstruction and Segmentation tasks}

After the intermediate feature $F_t$ is fused by the current-slice latent decoding $M_x$ and the auxiliary inter-slice latent decoding $M_F$, they can be fed into human-machine vision tasks like reconstruction and segmentation. Firstly, for restoring the images, we input the $F_t$ into a reconstruction network, which consists of three convolutional layers with kernel size $3\times3$. For the downstream segmentation tasks, we follow TransUnet\cite{transunet} to use Transformer\cite{attention} as the backbone to construct the image segmentation network. The TransUnet segmentation model employs a hybrid architecture that utilizes the pre-trained ResNet-50 and ViT\cite{vit} models as encoders while training a multi-upsampling decoder to produce the output masks. In our framework, we select the `R50-ViT-B16' hybrid pre-trained model. We also modify the channel number of the input layer to adapt to our latent decoding features $M_x$.

In this way, we achieve the target of using a single bitstream for both high-quality reconstruction tasks and machine vision segmentation tasks without decoding the coded representations into pixels. This kind of characteristic helps to facilitate the usage of our framework in digital healthcare scenarios such as medical clouds.

\section{Experiments}
In this section, we first introduce the datasets and the experimental settings. Then, we compare our method with the existing methods for both image reconstruction tasks and segmentation tasks on a variety of datasets. Finally, the ablation study and further analysis of our framework are provided.

\subsection{Datasets}

\subsubsection{MRNet}
The MRNet dataset consists of knee MRI examinations conducted on $1199$ patients, and $1370$ knee examinations\cite{mrnet}. Each examination includes three scanning orientations: axial, coronal, and sagittal. Each sequence of volumes consists of $17$ to $61$ slices that are stored as 8-bit $256 \times 256$ pixels. The training set includes $1130$ scans from $1088$ patients, while the validation set comprises $120$ scans from $113$ patients.

\subsubsection{CHAOS}
The CHAOS dataset\cite{chaos} aims at the segmentation of abdominal organs from CT and MRI data obtained by $40$ patients. We choose the MRI T2 dataset in our study, which supports multi-organ segmentation (i.e. liver, right kidney, left kidney, and spleen) tasks. Initially, these images are stored in $16$-bit Digital Imaging and Communications in Medicine (DICOM) format, the resolution of which ranges from $256 \times 256$ to $320 \times 320$ pixels. We arrange the patients in ascending order and divide the images into a training set ($14$ patients) and a validation set ($6$ patients). We also resize the images to $256 \times 256$ pixels through interpolation with the order equal to 3.

\subsubsection{ACDC}
The ACDC dataset\cite{ACDC} contains cardiac cine-MRI sequences obtained from $150$ patients. We directly use the divided training ($200$ sequences) and test sets ($100$ sequences) for abnormal organ segmentation tasks (i.e. left ventricle (LV), right ventricle (RV), and myocardium (MYO)). The raw input images are provided through the $16$-bit nifti format, the resolution of which ranges from $(154\sim428) \times (154\sim512)$. Similar to the CHAOS dataset, we resize the images to $224 \times 224$ pixels.

\subsubsection{Synapse}
We use the $30$ abdominal CT scans in the MICCAI 2015 Multi-Atlas Abdomen Labeling Challenge\footnote{\url{https://www.synapse.org/Synapse:syn3193805/wiki/217789}}, with $3779$ axial contrast-enhanced abdominal clinical CT images in total. Each CT volume consists of $85\sim198$ slices of $512\times512$ pixels. Following previous methods\cite{transunet, rahman2024multi}, we divide the datasets into a training set ($18$ volumes) and a testing set ($12$ volumes), utilizing $8$ abdominal organs, such as the aorta, gallbladder (GB), left kidney (KL), right kidney (KR), liver, pancreas (PC), spleen (SP), and stomach (SM).

\subsection{Experiment settings}
\subsubsection{Metrics}

For human vision tasks, we perform image reconstruction experiments to validate the effectiveness of our proposed framework, using Peak Signal-Noise Ratio (PSNR) and Bjøntegaard-delta metrics (BD-PSNR and BD-Rate) as the evaluation metrics for compression performance. For the segmentation machine vision task, following\cite{transunet, rahman2024multi}, we use DICE and $95\%$ Hausdorff Distance (HD95). 

\subsubsection{Baseline}
We conduct comparative experiments with traditional compression methods such as JP2K\cite{JP2K}, JPEG-XL\cite{jpegxl}, H.265/HEVC\cite{HEVC}, and VVC\cite{vvc}, as baseline methods. For JP2K\cite{JP2K}, we use the OpenJPEG-2.3.1 with the default configuration. For JPEGXL\cite{jpegxl} we utilize the software reference implementation of v0.10.2. We also utilize the FFmpeg-4.3 for compressing images with HEVC. As for VVC, we use the VTM-23.3 software with the $encoder\_lowdelay\_p\_vtm$ configuration file. Regarding learning-based lossy image compression, we compare our method with Minnen2018\cite{minnen_joint_2018}, Cheng2020\cite{cheng2020}, and Zou2022\cite{stf}. It’s noteworthy that we extend the number of latent channels from default $192$ to $256$ in the Cheng2020 method.

For the downstream segmentation experiment, we select JP2K as the anchor method due to its widespread use in medical image compression.

\subsubsection{Training Setting}

We adopt a unified network architecture for all experiments and set the channel depth of our auxiliary inter-slice feature $F$ to $16$. Then, we set the number of channels to 192 for both the latent and the hyper-latent features.

Typically, the training method for recurrent neural networks involves the Back Propagation Through Time (BPTT) algorithm\cite{BPTT}. However, when dealing with long-time sequences, the issue of vanishing gradients\cite{vanish} arises, corresponding to a large number of slices per volume in our datasets. To address these issues, we set the update stride as $16$. What’s more, during the inference process, this operation plays a similar role as the Group of Pictures (GOP) in traditional video-based compression methods. For each Group of sub-volumes, the first slice acts similarly to the I frame and the latter slices play the same role as the P frames.

First, we conduct image reconstruction experiments on three sub-datasets of MRNet. For rate-distortion optimization, we set the $\lambda$ values as follows: $\lambda=\{2048,4096,8192,16384\}$ for Axial dataset,  $\lambda=\{4096,8192,16384,32768\}$ for Coronal; and for Sagittal $\lambda=\{2048,4096,8192,12288\}$; To train our model on the MRNet datasets, we initialize the learning rate at $1\times10^{-4}$, set a total of $200$ epochs, and decay the learning rate by $80\%$ after every $20$ epoch. We employ the Adam optimizer with default settings in our experiments.

After that, to validate the performance of our method on segmentation datasets (including CHAOS, ACDC, and Synapse) for machine vision. Given that the size of these segmentation datasets is significantly smaller than that of MRNet, we adopt a fine-tuning approach to transfer the model trained on the MRNet dataset to the others. Specifically, we use MRNet-Axial ($\lambda=2048$) as the base model and fine-tune it for $100$ epochs with the same training settings. Subsequently, we utilize the retrained models for segmentation datasets to perform latent feature sampling on the raw images, obtaining current-slice latent decoding $M_x$, which is then fed into the segmentation network for analytics. 

\subsubsection{Platform}

During the training and inference phases, we use the NVIDIA RTX 3090 GPU and the Intel(R) Xeon(R) Gold 6128 CPU. Our code is implemented by Python 3.11 on PyTorch 2.0 framework.

\subsection{Reconstruction Results for Human Vision}

\begin{figure*}[!t]
\centering
\includegraphics[width=7in]{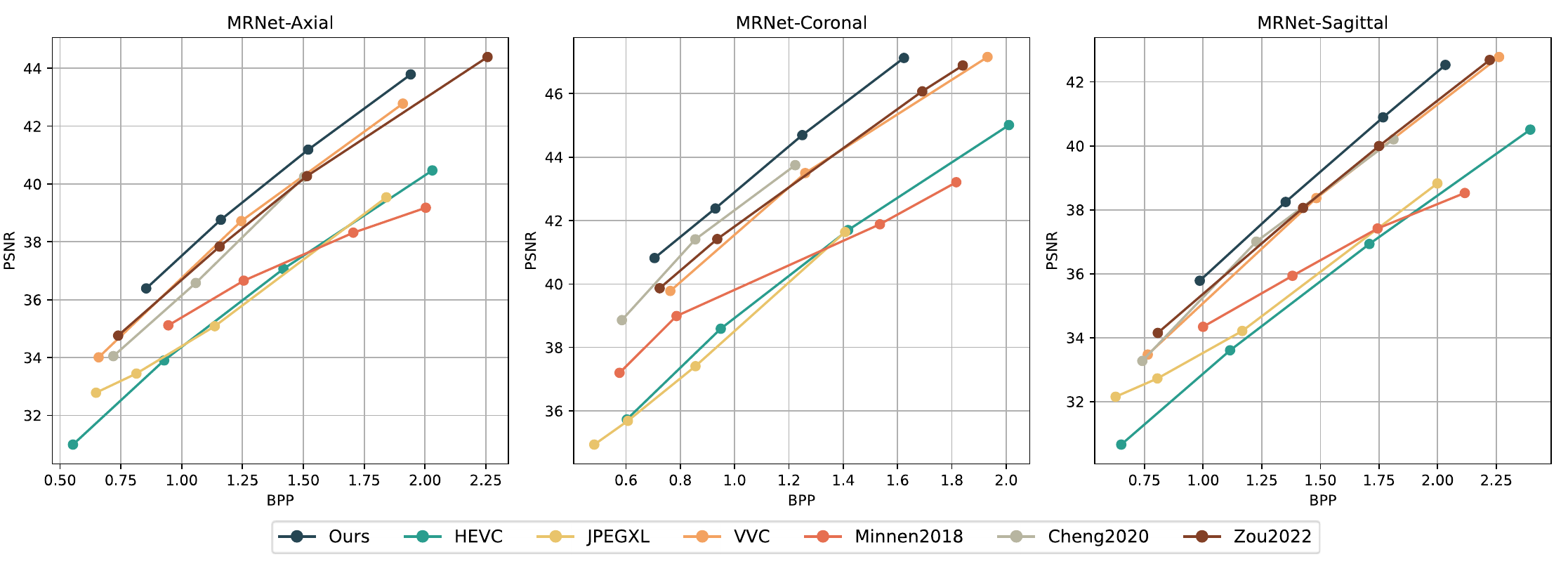}
\caption{Rate-distortion curves of various image compression approaches on MRNet datasets.}
\label{fig:mrnet}
\end{figure*}

\begin{table}[!t]
\renewcommand{\arraystretch}{1.1}
\setlength{\tabcolsep}{16pt}
\caption{BD-Rate ($\%$) Comparison for PSNR of Our Method and Other State-of-the-Art Cordecs on the MRNet Dataset}
\centering
\begin{tabular}{c|ccc}
\hline
Methods                      & Axial & Coronal & Sagittal   \\ \hline
VVC\cite{vvc}                & 0     & 0      & 0           \\
HEVC\cite{HEVC}              & 28.2  & 28.6   & 26.71       \\
JPEGXL\cite{jpegxl}          & 31.48 & 33.63  & 28.33       \\
Joint2018\cite{minnen_joint_2018} & 28.83 & 31.04  & 23.89  \\
Cheng2020\cite{cheng2020}    & 8.13  & -10.98 & -1.67       \\
Zou2022\cite{stf}                & 2.19  & -1.3   & -0.92       \\ \hline
VVMIC (ours)                         & -7.67 & -17.23  & -8.84      \\ \hline                 
\end{tabular}
\label{tab:mrnet_bdrate}
\end{table}

For the human vision-oriented dataset, MRNet\cite{mrnet}, we illustrate the rate-distortion curves in \figref{fig:mrnet} and list the detailed BD-Rate results in \tabref{tab:mrnet_bdrate}. When calculating the BD-Rate, we regard VVC\cite{vvc} as the anchor. Experimental results show that our proposed framework achieves an average 7.67\% BD-Rate reduction on the Axial sub-dataset against VVC in terms of PSNR. In the Coronal and Sagittal sub-datasets, our method achieves 17.23\% and 8.84\% BD-Rate improvements respectively compared to the VVC anchor. As for learning-based methods, our process also has 9.86\%, 18.53\%, and 9.76\% BD-Rate reductions than Zou2022\cite{stf} in all the three sub-datasets.

\begin{figure*}[!t]
\centering
\includegraphics[width=7in]{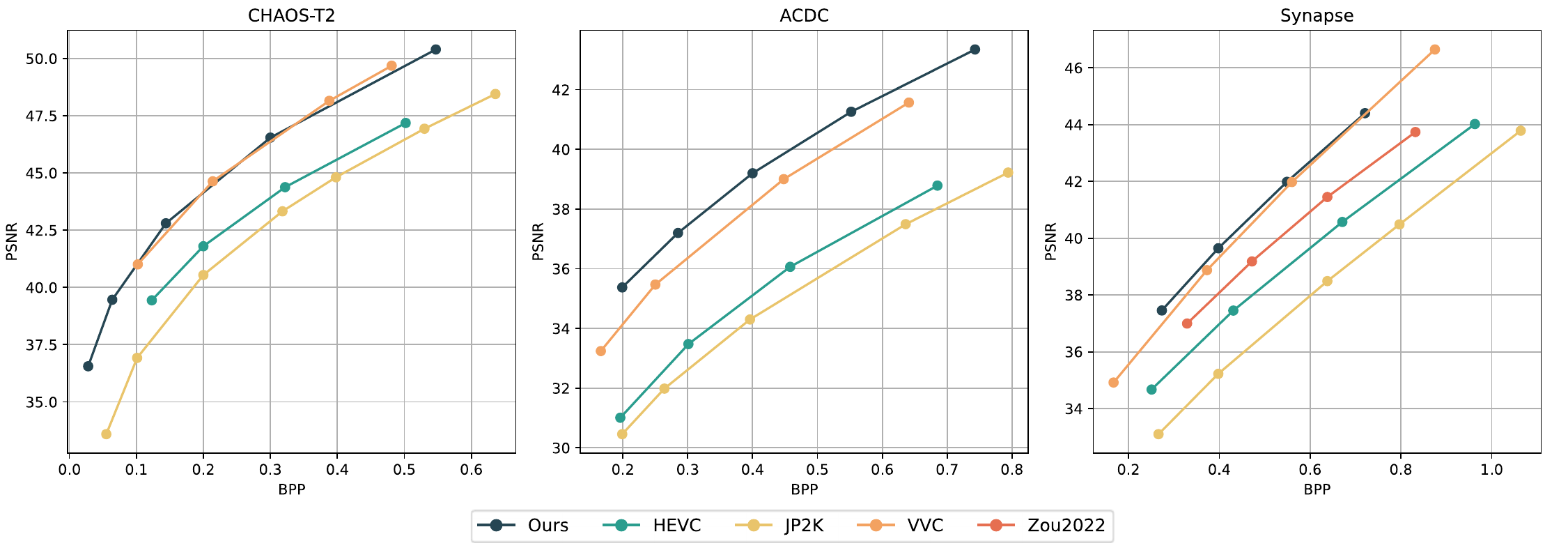}
\caption{Rate-distortion curves of various image compression approaches on the CHAOS, ACDC, and Synapse datasets.}
\label{fig:seg_compress}
\end{figure*}

For other machine vision-oriented datasets, including CHAOS\cite{chaos}, ACDC\cite{ACDC}, and Synapse, we also perform image reconstruction experiments. However, given that these datasets are relatively small, we employ the transfer learning approach to enhance the training efficiency. Additionally, considering that some compression networks\cite{cheng2020, stf} and the downstream segmentation network\cite{transunet} contain attention modules, which often require consistent input image dimensions, we perform resizing pre-processes on the CHAOS and ACDC datasets. To this end, we only apply some of the learning-based methods for comparison. The final rate-distortion curves are illustrated as \figref{fig:seg_compress}.

\begin{figure}[!t]
\centering
\includegraphics[width=3.4in]{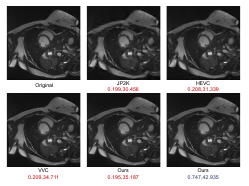}
\caption{Comparison of reconstruction quality for human vision on the ACDC dataset with compression results (BPP, PSNR).}
\label{fig:reconstruct}
\end{figure}

\subsection{Segmentatioon Results for Machine Vision}

In addition to human vision tasks, we further explore the effectiveness of using decoded intermediate features to perform machine vision tasks. We design two models Our Latents and Our Image. For Our Latents model, we feed the partially decoded intermediate features $M_x$ into the medical image segmentation networks shown in \figref{fig:framework}. For Our Image model, we first perform image reconstruction and then input the recovered pixel-level frames into the segmentation network. As for comparison, following\cite{liu2019machine}, we perform additional experiments using the traditional image compression algorithm JP2K. We first apply the JP2K algorithm to compress the CHAOS, ACDC, and Synapse datasets at compression rates of {20, 25, 40, 60, 80}. Subsequently, we feed the compressed images into the segmentation network to obtain the predicted masks. For Our Latents method, we adjust the number of input channels in the TransUnet network to fit the latent decoding feature. In Our Image and JP2K experiments, we modify the input channel number to $1$.

\begin{figure*}[!t]
\centering
\includegraphics[width=7in]{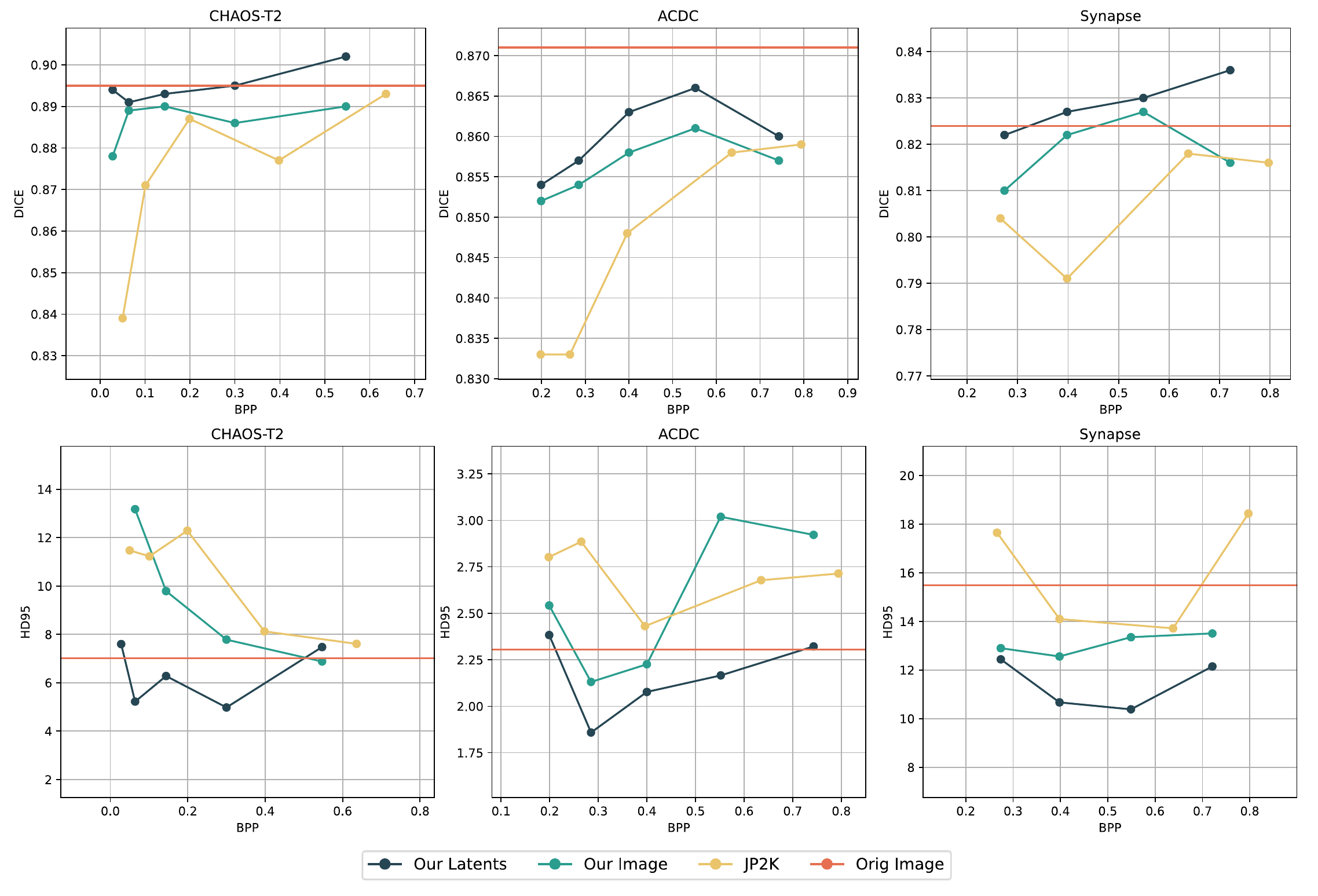}
\caption{Rate-performance (DICE and HD95 of the segmentation task) curves on the CHAOS, ACDC, and Synapse datasets, where higher DICE and lower HD95 metrics are better. Our Latents means using intermediate latent features as input, while Our Image uses fully decoded pixels.}
\label{fig:segmentation}
\end{figure*}

We present the detailed experimental results in \figref{fig:segmentation}. The red lines represent inputting the original images without compression into the TransUnet network, which means the bits per pixel are equal to the original data, i.e. 16. To compare with it, we perform experiments of three compression methods in different ratios and plot the DICE-BPP and HD95-BPP curves. 

The DICE coefficient is commonly used to calculate the similarity between ground truth and prediction in medical image segmentation. The Our Latents method demonstrates an improvement in DICE performance in our experiments. For example, as the bits per pixel (BPP) increases in the CHAOS dataset, the DICE performance of all three compression methods shows an upward trend. At the same compression rate, the Our Latents method achieves higher DICE scores than both the Our Image method and JP2K. Similarly, on the Synapse dataset, the DICE scores achieved by the Our Latents method are higher than those of JP2K across multiple BPP values. Notably, as illustrated in \figref{fig:segmentation}, the Our Latents method significantly improves segmentation performance at low BPP. We conduct further analysis on this phenomenon. From the reconstructed images shown in \figref{fig:reconstruct}, it can be observed that our method, based on autoencoders, produces clearer edge information under high compression ratios and low BPP situations. This greatly aids downstream segmentation networks in extracting organ edges, resulting in more accurate mask predictions. Additionally, reconstructing images clearly at low BPP is a significant challenge for the reconstruction network. Suppose our latent decoding features already contain sufficient information about the original images, but the reconstruction network is limited in its ability to recover a clear image, which explains why the DICE score of Our Latents method surpasses that of the Our Image method.

As for the HD95 metric, which is based on the calculation of the 95th percentile of the distances between boundary points in ground truth and prediction, it is often used to verify boundary errors in the worst cases. This means that a lower HD95 score indicates higher accuracy. \figref{fig:segmentation} shows the final experimental results and demonstrates that our Latents model achieves performance improvements across multiple datasets. For example, in the CHAOS dataset, the HD95 scores obtained under three different compression methods show a downward trend as the compression ratio decreases and the BPP increases. The Our Latents model achieves lower HD scores, even outperforming the original images under some BPP conditions. The experimental results on the Synapse dataset also confirm this. 

\begin{figure*}[!t]
\centering
\includegraphics[width=7in]{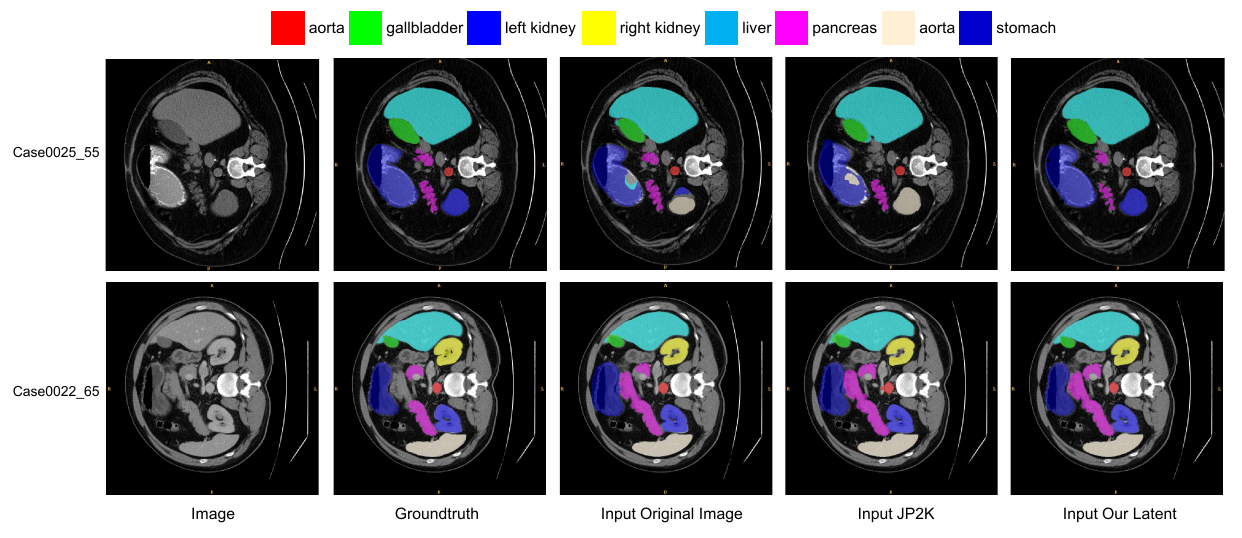}
\caption{Qualitative comparison of different segmentation approaches by visualization. From left to right: (1)original image (2)ground truth, (3)take the original image as input, (4)take the compressed image by JP2K as input, (5)take the latent decoding by our framework as input. In detail, the bits per pixel (BPP) of (4)JP2K is 0.266, and the BPP of (5)Our Latent is 0.2735.}
\label{fig:segmentation_visual}
\end{figure*}

\begin{table*}[!t]
\renewcommand{\arraystretch}{1.1}
\setlength{\tabcolsep}{14pt}
\centering
\caption{Ablation Results of Different Modules for Image Reconstruction task on MRNet-Axial dataset}
\begin{tabular}{c|ccccc}
\hline
 & $\lambda$ & BPP   & PSNR   & BD-Rate ($\%$) & BD-PSNR (dB) \\ 
 \hline
VVMIC (ours)                  & 2048  & 0.854 & 36.39  & 0           & 0                                \\
VVMIC - ckbd             & 4096  & 1.161 & 38.765 & 1.02        & 0.09                             \\
VVMIC - auxiliary        & 8192  & 1.520 & 41.19  & 3.06        & 0.27                             \\
VVMIC - channelwise      & 16384 & 1.941 & 43.792 & 3.57        & 0.32                             \\ \hline
\end{tabular}
\label{tab:ablation}
\end{table*}
\subsection{Ablation Study and Analysis}
\subsubsection{Analysis of Image Reconstruction Performance}
To further validate the impact of each modules in our framework on the compression performance of medical images, we modify several variants and retrain them on the MRNet-Axial dataset. First, we remove the checkerboard method from our Multi-dim Context Model, specifically eliminating the spatial context priors $\Psi_{sp}$ shown in \figref{fig:contextmodel}, and refer to this version as ‘ours - ckbd’. Similarly, we remove the channelwise context module $g_{ch}$ from our Multi-dim Context Model and name it ‘ours - channelwise’. Finally, to assess the impact of inter-slice auxiliary features, we remove the auxiliary analysis transform module $f_a$ and synthesis transform module $f_s$, calling this version ‘ours - auxiliary’. We obtain PSNR and bpp results under four different rate-distortion optimizations and calculate the BD-Rate (PSNR) metric, as shown in \tabref{tab:ablation}. It demonstrates that removing any one module leads to a decline in the final compression performance.

\subsubsection{Analysis of Different Latent Decoding for Downstream Segmentation}
\begin{figure}[!t]
\centering
\includegraphics[width=3.4in]{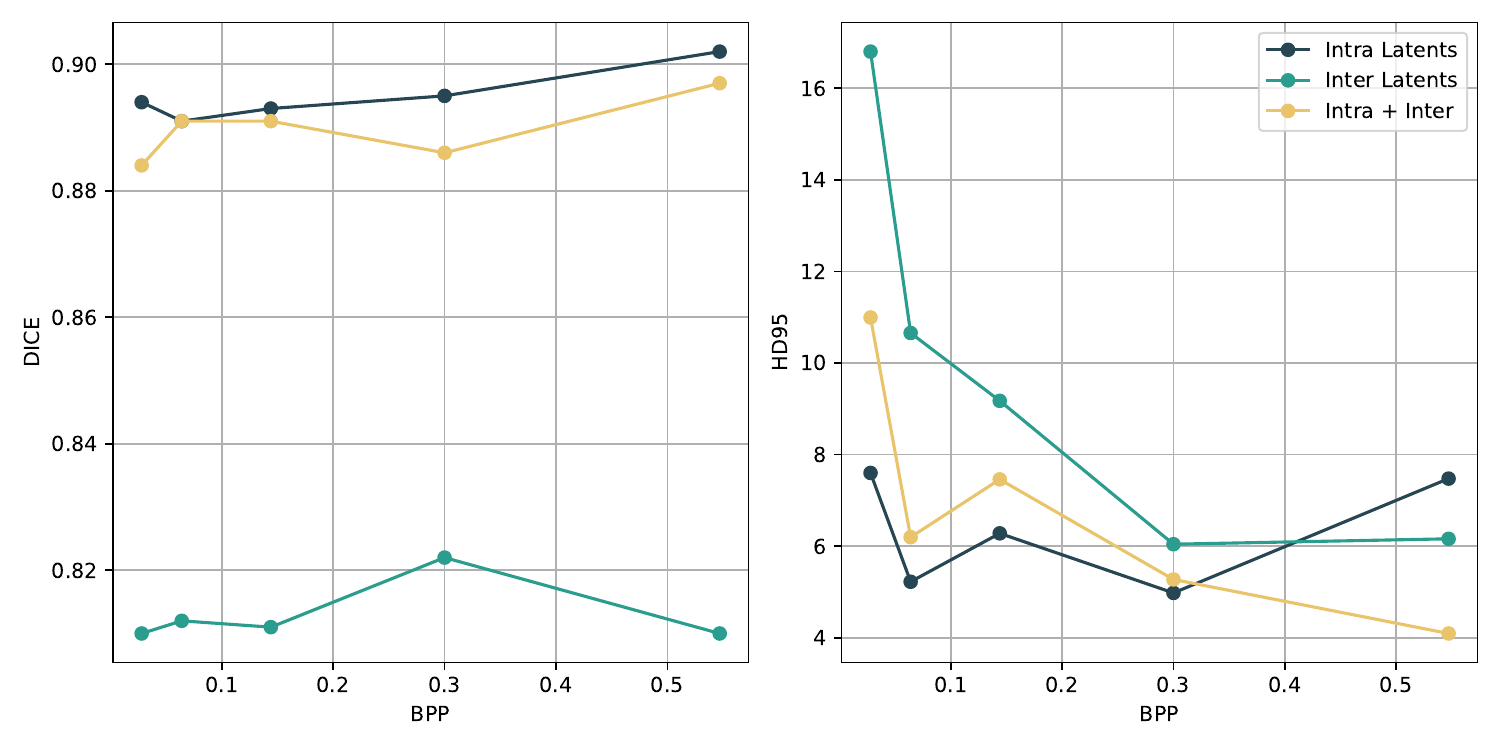}
\caption{Ablation study of different latent decoding features for downstream segmentation tasks on the CHAOS dataset.}
\label{fig:ablation_seg}
\end{figure}
Since We have mentioned in Section III and \figref{fig:framework} that we feed the current-slice latent decoding features $M_x$ into the downstream segmentation network, we perform further experiments to explore the impact of different latent features on the final segmentation performance. We first sample from the original images on the CHAOS dataset under different rates-distortion optimizations to obtain the current-slice latent decoding features $M_x$, the inter-slice latent decoding features $M_F$, and the updated inter-slice auxiliary feature $F_t$ respectively. Then, we retrain the TransUnet network with different decoding features as input. The final results are plotted in \figref{fig:ablation_seg}, which includes two sub-figures showing DICE-BPP and HD95-BPP curves. In most cases, the Intra Latents method represents current-slice latent decoding features $M_x$, outperforming the `Inter Latents' with $M_F$. Especially for the DICE metric, the `Intra Latents' method improves around 10 percent higher than the `Inter Latents' method. On the other hand, the `Intra + Inter' method represents the updated inter-slice auxiliary feature $F_t$, and achieves comparable results as the `Intra Latents' method but slightly worse. Thus in our segmentation tasks, we choose the current-slice latent decoding features $M_x$ as input for the TransUnet segmentation network.

\subsubsection{Analysis of Compression Complexities}
\begin{table}[!t]
\renewcommand{\arraystretch}{1.1}
\setlength{\tabcolsep}{6pt}
\centering
\caption{Average Encoding and Decoding Time per Image (in Seconds) on the MRNet-Axial Dataset}

\begin{tabular}{c|ccc}
\hline
                   & Encoding (s) & Decoding (s) & BD-Rate ($\%$) \\ \hline
VVC\cite{vvc}               & 24.062       & 0.014        & 0           \\
HEVC\cite{HEVC}               & 0.697        & 0.069        & 28.2        \\
JPEG-XL\cite{jpegxl}            & 0.061        & 0.031        & 31.48       \\
Zou2022\cite{stf}            & 0.071        & 0.096        & 2.19        \\ \hline
VVMIC - ckbd        & 0.064    & 0.085        & -6.21       \\
VVMIC - auxiliary   & 0.092    & 0.134        & -4.35       \\
VVMIC - channelwise & 0.041    & 0.046        & -3.49       \\
VVMIC (ours)               & 0.097        & 0.136        & -7.67       \\ \hline

\end{tabular}
\label{tab:speed}
\end{table}
In order to analyze the computational complexity of our framework, we compare the compression performance on MRNet Coronal dataset with traditional methods, including JPEG-XL\cite{jpegxl}, HEVC\cite{HEVC}, VVC\cite{vvc}, and the deep learning method Zou2022\cite{stf}, as well as the above-mentioned variants of our framework. The experimental results are shown in \tabref{tab:speed}. In traditional methods, VVC achieves the best compression performance, showing an improvement of $28.2\%$ BD-Rate over the HEVC and an improvement of $31.47\%$ over the JPEG-XL respectively. However, VVC's major drawback is its long encoding time, which limits its applicability for tasks sensitive to encoding latency. In contrast, learnable deep learning methods have similar encoding and decoding times and can achieve superior BD-Rate results than traditional compression methods. Our method saves $7.67\%$ in BD-Rate compared to the VVC standard while ensuring an acceptable coding speed.

\subsubsection{Analysis of Framework Limitations}
Our model simultaneously addresses both human vision-oriented image reconstruction tasks and machine vision-based image segmentation tasks. In terms of image reconstruction, our method outperforms traditional compression algorithms such as VVC, HEVC, and JPEG-XL, and shows a significant improvement in compression performance compared to learnable methods. The primary reason for this is that our approach is specifically designed for volumetric medical image data, proposing analysis and synthesis modules for auxiliary inter-slice features. We fuse the extracted inter-slice latent features into the current image features, thereby assisting the network in completing image reconstruction. However, our method still has limitations in extracting inter-slice latent features. For example, in our experiments, we set the update stride for inter-slice auxiliary features to 16, which means that this auxiliary feature is reset every 16 slices. In traditional video coding methods, the temporal context is extracted between the reference frame and the current frame, whereas our method extracts information from all previous frames. On the other hand, our framework is optimized for the entire dataset, meaning that it cannot achieve optimal rate-distortion optimization for individual volume data. Additionally, when there are slices with significantly varying ranges among different volumes in a dataset, it is challenging to extract the optimal inter-slice auxiliary priors. 

Furthermore, for machine-vision-based image segmentation tasks, our method uses the same bitstream as for image reconstruction, utilizing intermediate decoding features as input for the segmentation network, which saves time and resources needed to reconstruct pixel-level images. However, beyond image segmentation, machine vision downstream tasks also include image super-resolution, image denoising, image classification, object detection, and so on. Theoretically, the latent decoding features extracted by our framework contain expressive intra-slice and inter-slice information, which can assist downstream networks in various analytic tasks. We will conduct subsequent research and experiments in the future.

\section{Conclusion} 
This paper introduces a versatile volumetric medical image coding framework VVMIC for both human and machine vision. Since inter-slice redundancy between adjacent volumetric slices often arises from anatomical progressions through the body, we propose a recurrent-based VVAE module to incorporate the inter-slice features for the current-slice compression. Specifically, the auxiliary analysis and synthesis modules extract multi-scale inter-slice contexts, which are then concatenated to enhance the expressiveness of the current slice's latent features and aggregated in the multi-dimensional context model to improve the coding performance. 
The experimental results on four datasets demonstrate that our VVMIC framework outperforms traditional compression methods and existing learning-based networks for image reconstruction tasks. Moreover, using the coded representations of original images directly, our VVMIC achieves remarkable segmentation improvements over JP2K at various compression rates. Besides segmentation, this proposed framework can be easily tailored for other machine-vision applications, such as classification and detection, in digital healthcare scenarios.

\bibliographystyle{ieeetr}
\bibliography{ref}

\end{document}